\documentclass[aps,prstab,superscriptaddress,longbibliography, twocolumn,amsmath,amssymb]{revtex4-2}
\usepackage{graphicx}
\usepackage{dcolumn}
\usepackage{hyperref}

\hypersetup{
    colorlinks=true,
    linkcolor=blue,
    citecolor=blue,
    filecolor=magenta,
    urlcolor=blue
}
\newcommand{\Figref}[1]{Fig.\ref{#1}}
\newcommand{\Tabref}[1]{Table~\ref{#1}}

\begin{document}

\title{Acceleration gradients in dielectric laser accelerators with triangular-shaped gratings}

\author{O.O. Svystunov}
\altaffiliation[Corresponding author ]{\\Email address: oleg.svistunov.98@gmail.com (O.O. Svystunov)}

\author{I.V. Beznosenko}
\author{A.V. Vasyliev}
\author{R.R. Kniaziev}
\author{G.V. Sotnikov}

\address{National Science Center Kharkiv Institute of Physics and Technology \\
1, Akademichna St., Kharkiv, 61108, Ukraine}

\date{\today}

\begin{abstract}
The study investigated transparent on-chip structures with a rectangular profile and triangular profiles with grating ridge base angles of $\alpha = 36^\circ$, $30^\circ$, and $20^\circ$. Each triangular structure had both left- and right-handed profile orientations. For all variants, a modified version with a reflective gold coating was additionally considered. The maximum energy gains and accelerating gradients were determined and quantified for all structure classes: transparent and reflective (with both rectangular and triangular profiles).

\end{abstract}

\maketitle

\section{INTRODUCTION}
The development of charged particle acceleration technologies in recent decades has been focused on creating compact and highly efficient systems capable of providing high accelerating field gradients with minimal installation size. One of the most promising directions in this field is dielectric laser accelerators (DLAs), which are based on the interaction of optical radiation with micro- and nanostructured dielectric surfaces \cite{Sotnikov2025,Vasyliev2022,Bolshov2021,Andonian2025}. According to current research, such structures have the potential to achieve accelerating gradients on the order of several GeV/m \cite{Cesar2018}, opening prospects for the development of compact tabletop accelerators for fundamental science and applied technologies \cite{Cohen2020,Liu2025,Fernow1991}.

The relevance of DLA structure research is driven by several factors. Firstly, the rapid advancement of laser sources with femtosecond pulse duration and high peak power enables the excitation of intense modulated fields over periodic dielectric structures without causing their damage, which is a distinctive feature of these lasers \cite{Stuart1995,Lenzner1999}. Secondly, progress in fabrication technologies for such micro-scaled structures, including electron-beam lithography and reactive ion etching, allows for the creation of elements with sub-micron accuracy. These circumstances facilitate the practical implementation of complex on-chip structures with given geometry.

\begin{figure}[!bh]
  \centering
  \includegraphics[width=\textwidth]{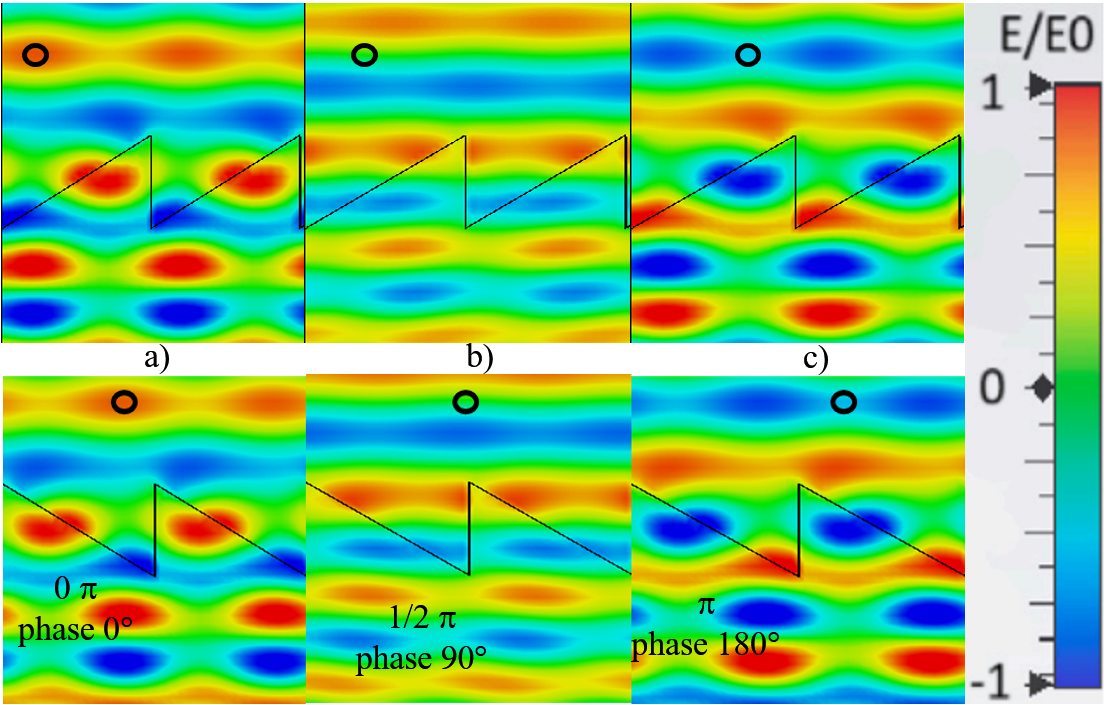}
  \caption{Illustration of the particle acceleration principle in triangular on-chip structures: accelerating (a), neutral (b), and decelerating (c) field regions near the dielectric structure surface. Here, $E_0$ represents the maximum axial electric field value, therefore warm colors ($E/E_0 > 0$) correspond to the accelerating phase of the electric field acting on the electron, while cool colors ($E/E_0 < 0$) represent the decelerating phase. The accelerated electron is marked by a black circle.}
  \label{Fig:01}
\end{figure}

The key operating principle of a DLA (\Figref{Fig:01}) is to ensure a regime where a charged particle passes over a grating ridge precisely at the moment when the laser pulse's electric field is accelerating (co-directional with the particle's motion) \cite{Breuer2013}. The surface profile of the dielectric structure determines the spatial distribution of the electric field and directly influences the acceleration efficiency. The greatest attention has traditionally been paid to rectangular and sinusoidal profiles of dielectric structures (see \cite{Sotnikov2025,Vasyliev2021} and references therein). However, among the multitude of available periodic structure options, triangular profiles are of interest for DLAs, as they should provide an increase in the acceleration rate compared to rectangular or sinusoidal structures. The fabrication technology for dielectric structures with triangular profiles of a periodic surface is well-developed, and such structures are produced industrially.

Despite the existence of publications devoted to the study of DLAs using periodic structures \cite{Sotnikov2025,Vasyliev2022}, DLAs with a triangular profile have not been studied in detail. In particular, a comparative analysis of triangular on-chip structures with different apex angles and profile orientations has not been conducted. It should be noted that the work \cite{Vasyliev2021} performed a comparison of an on-chip structure with a triangular profile and structures with different profiles. For the DLA with a triangular profile, the authors selected a structure with a single, fixed profile.

The acceleration of charged particles in periodic structures is based on the reversed Smith-Purcell effect \cite{Smith1953,Palmer1980}. As an illustration, \Figref{Fig:01} shows the motion of an electron in a triangular dielectric structure irradiated by a laser pulse. The electron is accelerated in the region of space where the particle is in an accelerating field, which propagating at an angle relative to the apex of the structure's triangular grating profile (\Figref{Fig:01}a). After a quarter of a period in the interaction space with the electron, a neutral field is established, corresponding to the moment when the electric field direction reverses (\Figref{Fig:01}b). After another quarter period, when the field polarity becomes opposite, the electron finds itself in a decelerating region (\Figref{Fig:01}c), which, due to the profile of the dielectric structure, has a lower intensity than the accelerating one. Consequently, the electron gains energy per structure period from its interaction with the field. The condition for phase synchronism is as follows \cite{Breuer2013}:
\begin{equation}\label{eq:01}
{\lambda _p} = n\beta {\lambda _L},
\end{equation}
where, $n$ is the mode of the electromagnetic field excited by the incident wave; $\beta  = v/c$ is the dimensionless velocity, $v$ is the electron velocity, $c$ is the speed of light; $\lambda_L$ is the wavelength of the driving laser radiation.

The aim of this work is to perform a numerical analysis of the influence of the apex angle and mirror orientation of triangular profile ridges on the electron accelerating gradient in dielectric on-chip structures, as well as to compare the resulting accelerating gradients with those for structures possessing a rectangular profile. Additionally, the influence of a reflective coating on the acceleration characteristics is investigated.

\onecolumngrid

\twocolumngrid

\section{PROBLEM FORMULATION}
\begin{figure}[!bh]
  \centering
  \includegraphics[width=\textwidth]{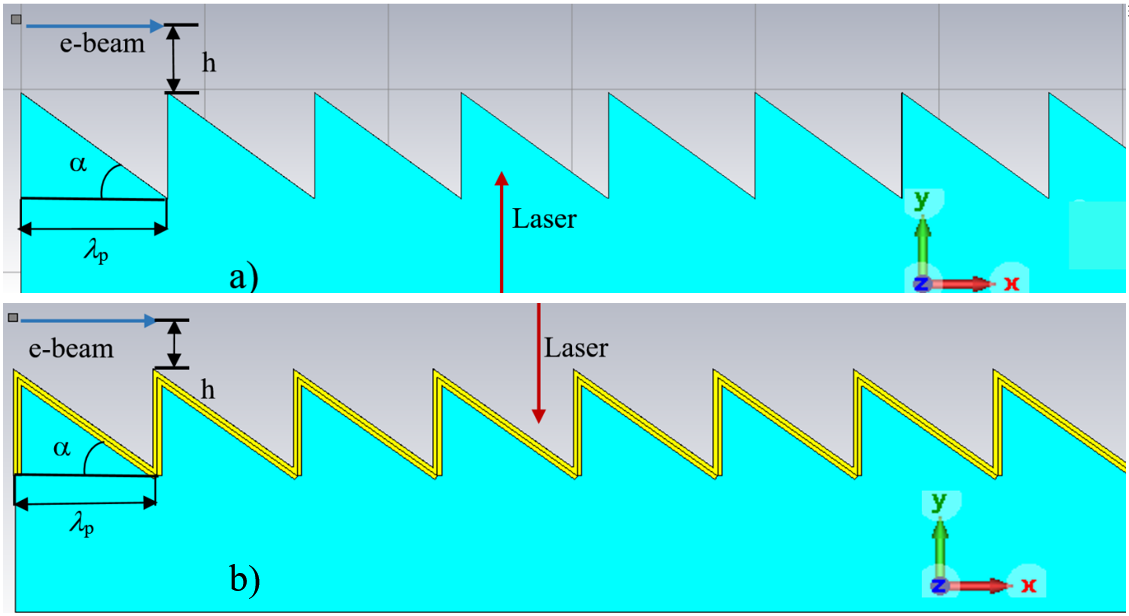}
  \caption{Scheme of the theoretical model for the numerical simulation of acceleration in a DLA with a triangular profile for transparent (a) dielectric and reflective (b) surfaces. This figure corresponds to the left-handed orientation of the triangular profile.}
  \label{Fig:02}
\end{figure}

The geometry of the triangular periodic structure is shown in \Figref{Fig:02}. The direction of the electron beam propagation is also indicated; the laser pulse propagates perpendicular to the structure's substrate in all investigated configurations. Numerical simulations were performed using a Particle-In-Cell (PIC) algorithm within CST Studio Suite \cite{CST2024}, which enables modeling the self-consistent interaction between the electron beam and the electromagnetic field excited by the laser radiation.

All investigated on-chip structures have a length of $L$= 20 $\mu$m and a grating period of $\lambda_p$ = 800 nm. The grating period corresponds closely to the wavelength of the driving laser pulse, $\lambda_L$ = 800 nm. Fused silica with a relative permittivity of $\varepsilon$ = 2.11 at the wavelength $\lambda$ = 800 nm was chosen as the material for the dielectric structure.

The following configurations were investigated:
\begin{itemize}
\item Triangular profiles with ridge base angles of $\alpha = 36^\circ$, $30^\circ$ and $20^\circ$;
\item Each triangular structure featured both left- and right-handed profile orientations. Herein, a left-handed orientation refers to a structure with ridges facing left, while a right-handed orientation refers to a structure with ridges facing right;
\item A rectangular profile (base case) with a pillar height and width of 400 nm and 400 nm, respectively, for the transparent structure, and 200 nm and 400 nm, respectively, for the reflective structure;
\item For all profile variants, a modified version with a gold coating (electrical conductivity $\sigma  = 45.6 \cdot {10^6}{\rm{ S/m}}$) applied to the structure was additionally considered.
\end{itemize}

The selection of the ridge base angles was motivated, firstly, by the availability of ready-made on-chip structures from manufacturers with similar base angle specifications (Thorlabs \cite{Thorlabs2024}, OpticElectronics), and secondly, by the interest in a comparative analysis of the influence of ridge orientation and profile angle on the accelerating gradient.

The laser pulse was approximated as a plane wave incident perpendicularly from the substrate side for transparent structures (\Figref{Fig:02}a). For reflective structures, the wave was injected from the grating ridges side, corresponding to the excitation scheme used in actual experiments (\Figref{Fig:02}b). The electric field strength of the wave was 1 GV/m. A direct current (DC) electron beam was used as the particle source. The initial electron energy was $E_{e0}$ = 10 MeV. The beam propagated along the surface of the on-chip structure, and its trajectory in the surface plane was aligned with the X-axis (\Figref{Fig:02}). The distance from the beam propagation axis to the structure surface was $h$ = 400 nm. Particles were injected from a cathode shaped as a square with 50 nm sides. Accordingly, the electron beam thickness was 50 nm.

It should be noted that this work does not account for effects related to the Gaussian distribution of the laser pulse or its temporal profile, which can significantly limit the maximum achievable gradients. Furthermore, the simulations were performed for single electrons and idealized beams. In a real experiment, space charge effects, emittance, and angular divergence can substantially influence the acceleration efficiency.

The primary objective of the simulation was to determine the electron energy gain $\Delta E$ after passing along the structure of length $L$, the accelerating gradient $G = \Delta E / L$, and the dependence of the acceleration efficiency on the angle $\alpha$, the mirror symmetry of the profile, and the presence of a reflective coating.

The use of a DC electron source model, instead of a model with short electron beams of finite duration, is justified by the need to determine the maximum possible electron energy gain. When using short beams (shorter than the wavelength) as an electron source, precise beam positioning relative to the wave phase is required to find the maximum achievable energy gain. Furthermore, this DC model ensures an objective comparison of the efficiency of different accelerating structures.

\onecolumngrid

\twocolumngrid

\section{TRANSPARENT STRUCTURES}
\subsection{Transparent rectangular structure}
\begin{figure}[!bh]
  \centering
  \includegraphics[width=0.5\textwidth]{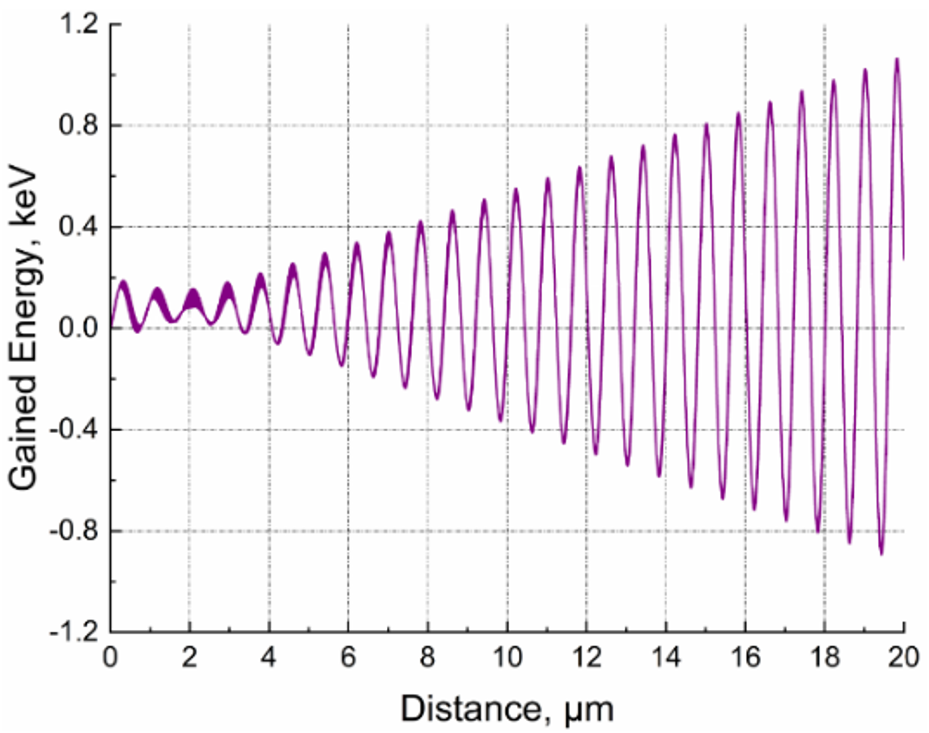}
  \caption{Electron beam phase space "energy gain vs. longitudinal coordinate" in a DLA with a rectangular on-chip structure.}
  \label{Fig:03}
\end{figure}

For further comparison of the energy gain values of electrons passing along on-chip structures with a triangular profile, the change in energy of the electron beam particles obtained in a DLA with a rectangular surface profile (\Figref{Fig:03}) was determined. This type of on-chip structure profile is widely used and well-studied in the literature \cite{DCesar2018,Mei2023}, therefore it was chosen as a reference sample.

The parameters of the incident wave and the structure period were identical to those of the model used for analyzing beam propagation with on-chip structures featuring a triangular profile.

\Figref{Fig:03} (and similar graphs for transparent and reflective structures) shows the energy distribution across the entire electron flux along the full length of the on-chip structure at a specific moment in time. Here, positive energy gain values represent electrons experiencing an accelerating phase, while negative values correspond to electrons in a decelerating phase.

As can be seen from \Figref{Fig:03}, when the beam propagates over the structure with a rectangular profile, the maximum electron energy gain is $\Delta E \approx 1.1$ keV, corresponding to an accelerating gradient of $G \approx 55$ MeV/m. This aligns with previously obtained results where rectangular and sinusoidal structures provide accelerating gradients in the range of 50 --150 MeV/m \cite{Sotnikov2025,Vasyliev2022,Vasyliev2021}.

\subsection{Transparent triangular structures}
\begin{figure}[!bh]
  \centering
  \includegraphics[width=0.96\textwidth]{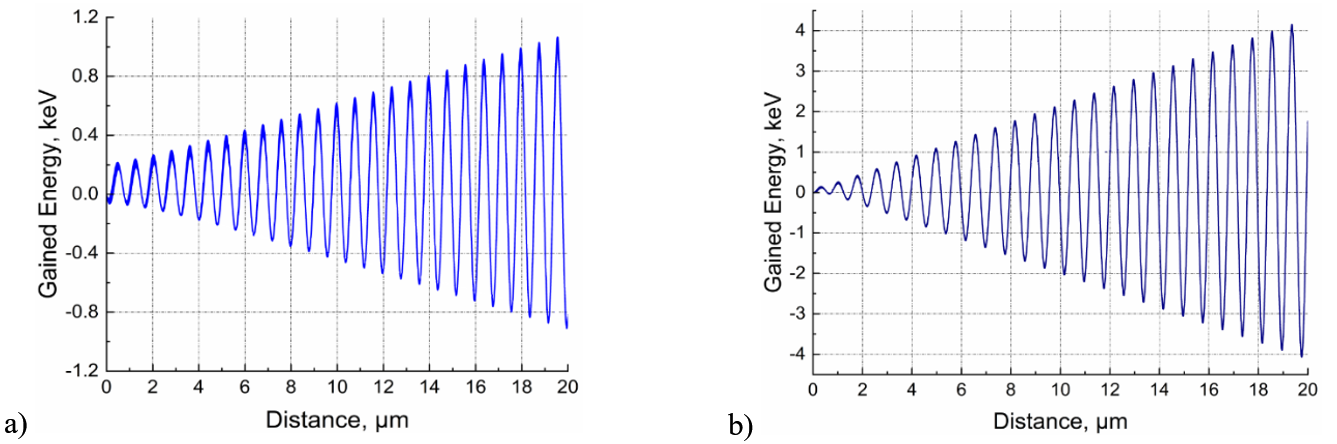}
  \caption{Electron beam phase space "energy gain vs. longitudinal coordinate" in DLAs with a left-handed (a) and right-handed (b) transparent ridge orientations and a ridge base angle of $\alpha = 36^\circ$.}
  \label{Fig:04}
\end{figure}

\Figref{Fig:04} presents the simulation results of electron beam acceleration in DLAs with triangular dielectric structure profiles with different ridge orientations and a ridge base angle of $\alpha = 36^\circ$. Quantitative values are provided in \Tabref{Tabl_1} below.

The maximum energy of accelerated electrons for both right-handed and left-handed on-chip structures was achieved with a ridge base angle of $\alpha = 36^\circ$. The energy gain for the transparent right-handed structure was $\Delta E \approx 4.1$ keV, which is four times higher than the result for the rectangular profile. For left-handed structures, the efficiency was significantly lower: even at the optimal angle of $\alpha = 36^\circ$, the energy gain was approximately 1.1 keV, equaling the energy gain obtained from the rectangular structure.
Thus, the analysis of the results reveals a substantial asymmetry in the efficiency of left- and right-handed structures. Furthermore, an additional study was conducted on right-handed transparent structures with ridge base angles of $\alpha = 40^\circ$ and $33^\circ$. The maximum energy gained for the structure with $\alpha = 40^\circ$ was $\Delta E \approx 2.3$ keV, while for the structure with $\alpha = 33^\circ$ it was $\Delta E \approx 1.8$ keV. This indicates that the result obtained with the right-handed on-chip structure with a ridge base angle of $\alpha = 36^\circ$ is not a simulation artifact, but rather a favorable configuration capable of delivering a high acceleration rate.

It is noteworthy that the minimum gained energy of the most accelerated electron among all investigated right-handed structures was 800 eV (at $\alpha = 20^\circ$). This result is nearly equal to the maximum acquired energy of the most accelerated electron obtained for the left-handed modifications.

Therefore, based on the analysis of the values presented in \Tabref{Tabl_1}, it can be concluded that the use of the investigated modifications of left-handed "transparent" structures is not promising, as the corresponding right-handed configurations demonstrate significantly higher electron acceleration efficiency.

\onecolumngrid

\twocolumngrid

The accelerating and decelerating electric fields of the laser radiation in transparent on-chip structures with a triangular profile do not propagate perpendicularly to the motion of the accelerated electrons. In the case of left-handed structures, the projection of these field vectors onto the X-axis (\Figref{Fig:01}, \Figref{Fig:02}a) does not coincide with the direction of electron motion, whereas in right-handed structures, it does. Consequently, when propagating over right-handed transparent triangular on-chip structures, the accelerated electrons remain in the accelerating phase of the field for a longer duration compared to similar left-handed transparent structures, where a rapid alternation between accelerating and decelerating phases occurs due to a counter-propagating phase front. Interference between beams refracted at different facets further reinforces this effect. Therefore, unlike left-handed transparent triangular on-chip structures and rectangular transparent on-chip structures, right-handed triangular transparent on-chip structures provide a higher acceleration gradient.

\begin{table}
   \centering
   \caption{Maximum energy gain and maximum accelerating gradient for electrons in DLAs with transparent on-chip structures}
   \begin{tabular}{lcc}
   \hline
   \hline
       \textbf{Structure type} & \textbf{Energy gain} & \textbf{Accelerating gradient} \\ &\hspace{15pt}\textbf{$\Delta E$, keV}  &\hspace{15pt}\textbf{$G$, MeV/m} \\
   \hline
   \hline
          Left-handed \\ transparent, $\alpha = 36^\circ$ & 1.1 & 55 \\
          Left-handed \\transparent, $\alpha = 30^\circ$ & 0.5 & 25 \\
          Left-handed \\transparent, $\alpha = 20^\circ$ & 0.47 & 23.5 \\
          Right-handed \\transparent, $\alpha = 36^\circ$ & 4.1 & 205 \\
          Right-handed \\transparent, $\alpha = 30^\circ$ & 1.5 & 75 \\
          Right-handed \\transparent, $\alpha = 20^\circ$ & 0.8 & 40 \\
          Rectangular \\transparent & 1.1 & 55 \\
   \hline
   \hline
   \end{tabular}
   \label{Tabl_1}
\end{table}

\section{REFLECTIVE STRUCTURES}
\subsection{Reflective rectangular structure}
\begin{figure}[!bh]
  \centering
  \includegraphics[width=0.5\textwidth]{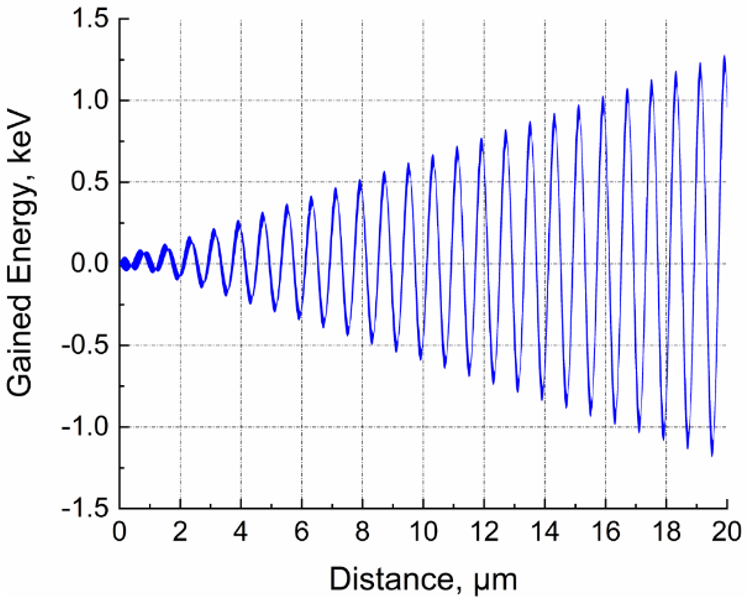}
  \caption{Electron beam phase space "energy gain vs. longitudinal coordinate" in a DLA with a reflective rectangular on-chip structure.}
  \label{Fig:05}
\end{figure}

Simulations of electron acceleration in a DLA with a reflective rectangular on-chip structure profile were also conducted. \Figref{Fig:05} shows the energy distribution of the electron beam along the accelerating structure at a fixed moment in time. All wave and beam parameters were identical to those used in the model for the on-chip structures with a triangular profile.

As can be seen from \Figref{Fig:05}, when propagating over the structure with a rectangular profile, the maximum electron energy gain is $\Delta E \approx 1.3$ keV, corresponding to an accelerating gradient of $G \approx 65$ MeV/m. Switching to a reflective rectangular on-chip structure yields a profit in the energy gain value of only $\sim$200 eV compared to its transparent counterpart. This makes the use of such a structure impractical, since the damage threshold of fused silica is approximately three times higher than that of gold. Therefore, further comparisons of the results obtained using reflective triangular structures will be made against the results from the transparent rectangular structure.

\subsection{Reflective triangular structures}
\begin{figure}[!bh]
  \centering
  \includegraphics[width=\textwidth]{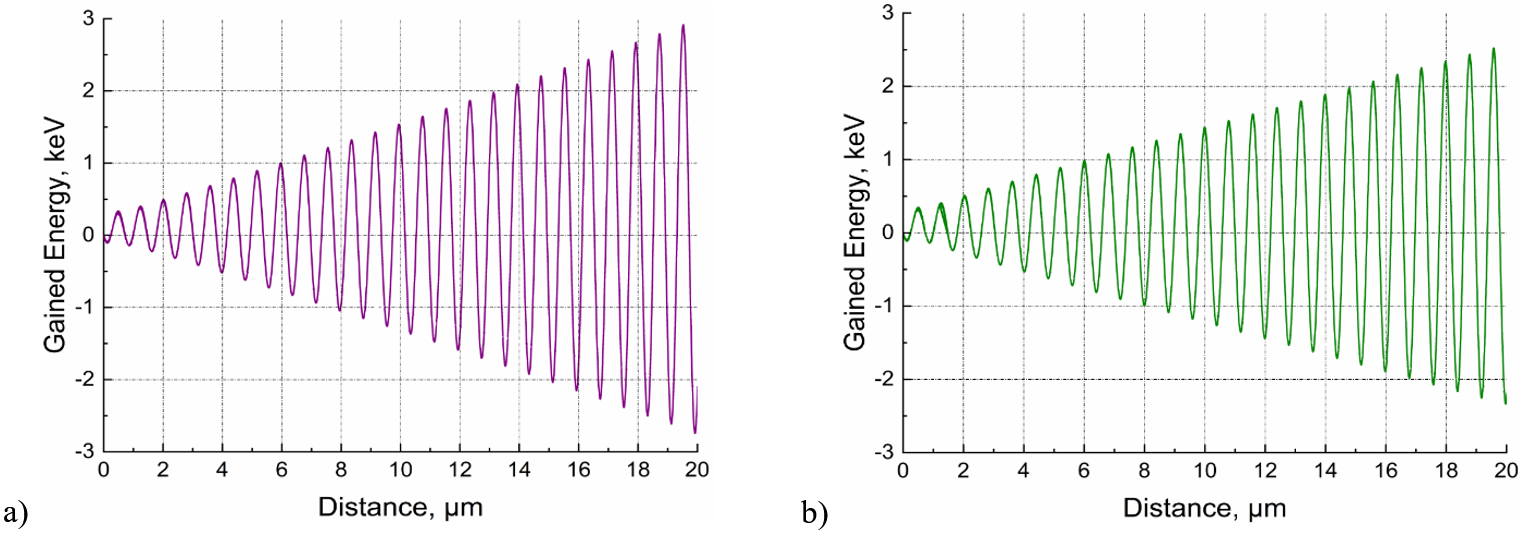}
  \caption{Electron beam phase space "energy gain vs. longitudinal coordinate" in DLAs with a left-handed (a) and right-handed (b) reflective ridge orientations and a ridge base angle of $\alpha = 20^\circ$.}
  \label{Fig:06}
\end{figure}

\Figref{Fig:06} presents the simulation results for the DLA with reflective triangular profiles featuring different ridge orientations and a ridge base angle of $\alpha = 20^\circ$. Quantitative values for the remaining structures are presented in \Tabref{Tabl_2} below.

Based on the conducted research, the maximum electron energy gained in reflective on-chip structures is observed at a ridge base angle of $\alpha = 20^\circ$.

For the left-handed configuration (\Figref{Fig:06}a), the maximum electron energy gain is $\Delta E \approx 3$ keV, which is approximately three times higher than the corresponding value for the rectangular structure. The right-handed structure provided a maximum electron energy gain of $\Delta E \approx 2.5$ keV. It is important to note that the efficiency of left-handed structures strongly depends on the ridge base angle $\alpha$: the acquired energy decreases to 2 keV at $\alpha = 36^\circ$.

\begin{table}
   \centering
   \caption{Maximum energy gain and maximum accelerating gradient for electrons in DLAs with reflective on-chip structures}
   \begin{tabular}{lcc}
   \hline
   \hline
      \textbf{Structure type} & \textbf{Energy gain} & \textbf{Accelerating gradient} \\ &\hspace{15pt}\textbf{$\Delta E$, keV}  &\hspace{15pt}\textbf{$G$, MeV/m} \\
   \hline
   \hline
          Left-handed \\reflective, $\alpha = 36^\circ$ & 2.0 & 100 \\
          Left-handed \\reflective, $\alpha = 30^\circ$ & 2.5 & 125 \\
          Left-handed \\reflective, $\alpha = 20^\circ$ & 3.0 & 150 \\
          Right-handed \\reflective, $\alpha = 36^\circ$ & 2.3 & 115 \\
          Right-handed \\reflective, $\alpha = 30^\circ$ & 2.4 & 120 \\
          Right-handed \\reflective, $\alpha = 20^\circ$ & 2.5 & 125 \\
          Rectangular \\reflective & 1.3 & 65 \\
   \hline
   \hline
   \end{tabular}
   \label{Tabl_2}
\end{table}

At the same time, right-handed reflective on-chip structures exhibit a weak dependence of the maximum electron energy gain on changes in the ridge base angle $\alpha$. The energy gain varies within a narrow range from 2.3 to 2.5 keV. In contrast, for DLAs with left-handed triangular structure profiles, the spread in energy gain values is significantly higher: ranging from 2 to 3 keV for the same change in angle from $36^\circ$ to $20^\circ$.

The resultant accelerating and decelerating electric fields of the wave in reflective on-chip structures with a triangular profile do not propagate perpendicularly to the motion of the accelerated electrons. For left-handed structures, the projection of these field vectors onto the X-axis (\Figref{Fig:02}b) coincides with the direction of electron motion, whereas for right-handed structures, it is opposite. In other words, when propagating over left-handed reflective triangular on-chip structures, the accelerated electrons remain in the accelerating field for a longer duration compared to similar right-handed reflective on-chip structures. The decelerating field encountered by the electrons has low intensity due to diffraction effects, as part of this field is reflected from the groove (the base of the ridge). Because such a groove has a sharp-angled profile, these diffraction effects are much stronger than when laser radiation is reflected from a groove with a rectangular profile. Therefore, unlike transparent on-chip structures, both left-handed and right-handed triangular reflective on-chip structures provide greater acceleration than rectangular reflective on-chip structures.

\onecolumngrid

\twocolumngrid

\section{ACCELERATION OF POINT BUNCHES}
In previous sections, our investigations of transport of a continuous beam along on-chip structures irradiated by a laser pulse are described. In this formulation, the output of the structure contains both accelerated and decelerated particles. Real accelerators utilize short particle bunches to "capture" the maximum number of electrons in the acceleration process. This requires tuning the bunch injection delay time relative to the laser pulse injection.

\begin{figure}[!bh]
  \centering
  \includegraphics[width=0.5\textwidth]{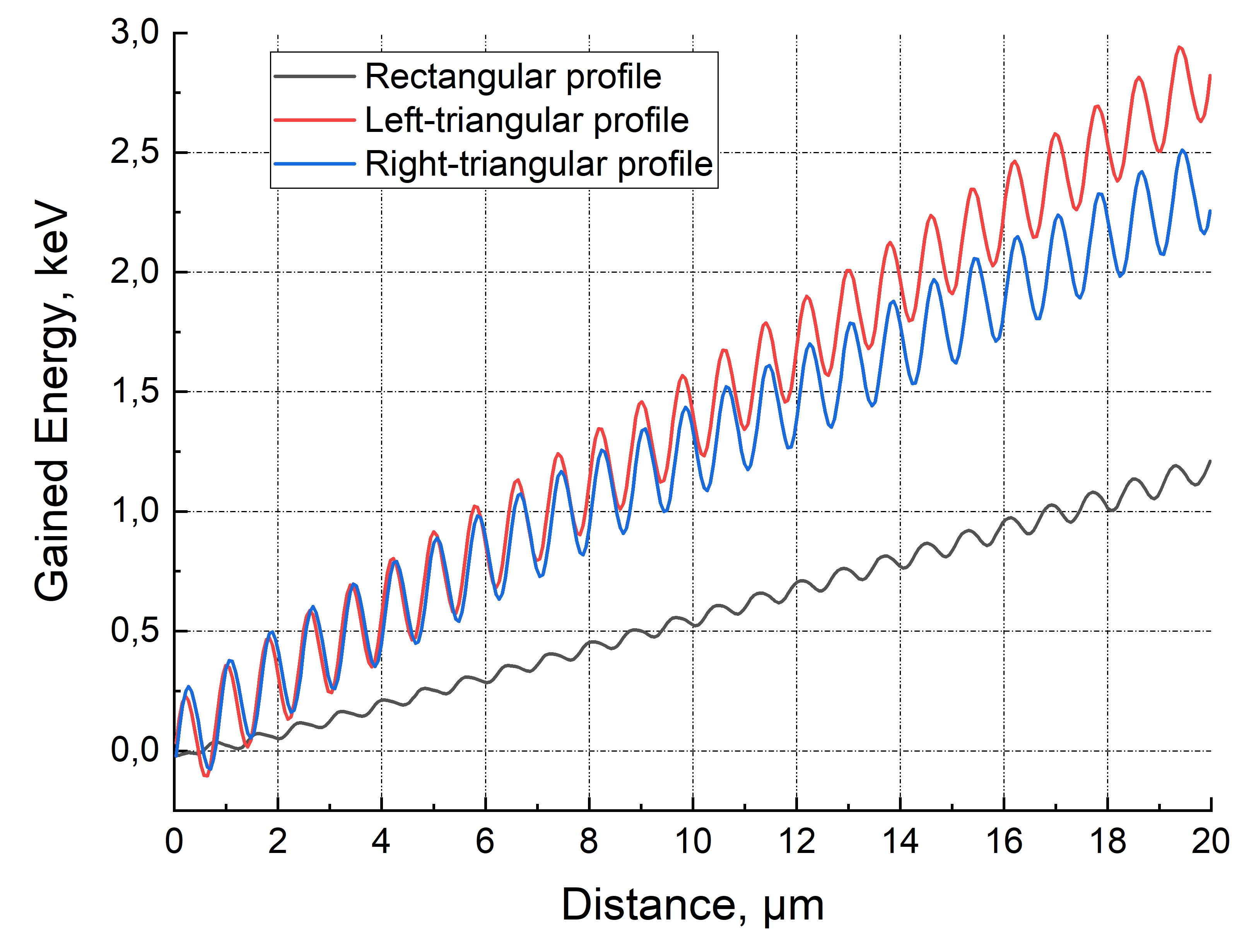}
  \caption{Change in electron energy along on-chip structures in DLAs at the optimal injection time relative to the wave phase for a rectangular reflective structure, a left-handed reflective structure with a ridge base angle of $\alpha = 20^\circ$, and a right-handed reflective structure with a ridge base angle of $\alpha = 20^\circ$.}
  \label{Fig:07}
\end{figure}

In this section, our investigations of the acceleration process of a point-like bunch (a single electron) as a function of its delay time relative to the laser pulse are described. We also examined the dependence of the electron energy change on the distance traveled when the injection time is optimized to yield the maximum electron energy gain at the structure's output. This analysis is performed for the following structures: a reflective rectangular structure, a left-handed reflective structure with a ridge base angle of $\alpha = 20^\circ$, and a right-handed reflective structure with a ridge base angle of $\alpha = 20^\circ$ (\Figref{Fig:07}).

Analysis of the data presented in \Figref{Fig:07} shows that the obtained electron energy gain values fall slightly short of the maxima reported in \Figref{Fig:05} and \Figref{Fig:06}. For the structure with a rectangular profile, the injection time yielding the maximum energy gain of 1.22 keV was 0.15 fs. For the reflective left-handed on-chip structure, it was 1.4 fs with an energy gain of 2.48 keV, and for the reflective right-handed structure, it was 1.45 fs with an energy gain of 2.95 keV. This discrepancy is attributed to differences in the simulation's initial conditions.

In the study presented in \Figref{Fig:07}, the electron was injected precisely from the geometric center of the electron gun's emitting surface. In contrast, for the data in \Figref{Fig:05} and \Figref{Fig:06}, electron emission from the entire cathode area was modeled, resulting in a beam with a wide spread of initial electron coordinates. In the latter case, some electrons propagate at a lower height above the structures, where the field gradient is higher than along the axis of the cathode's geometric center. This explains the registration of higher energy values.

\begin{figure}[!bh]
  \centering
  \includegraphics[width=0.5\textwidth]{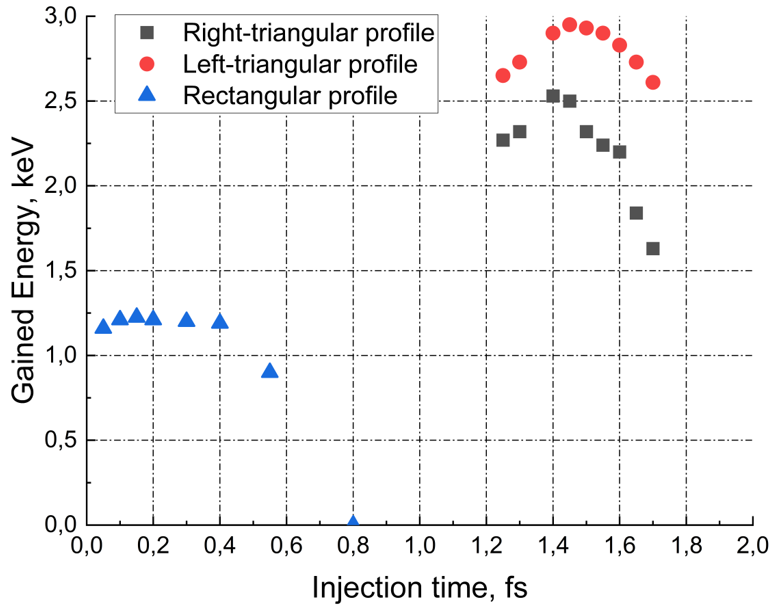}
  \caption{Dependence of electron energy on injection time for various reflective on-chip structures.}
  \label{Fig:08}
\end{figure}

Based on the study of the acceleration dynamics of a single electron propagating over reflective rectangular, left-handed reflective triangular, and right-handed reflective triangular on-chip structures, the dependence of the final electron energy on its injection time was analyzed. The corresponding plots are shown in \Figref{Fig:08}. The data points were selected with a step of 0.05 fs from the injection time that yielded the maximum energy gain.

Analysis of \Figref{Fig:08} reveals that for the right-handed reflective structure, the injection time yielding the maximum energy gain is 1.4 fs, whereas for the left-handed reflective structure, it is 1.45 fs. For the reflective rectangular on-chip structure, this time is 0.15 fs. This spread in values is a direct consequence of the distinct dynamics of electromagnetic field formation and interaction in each chip structure configuration. The close, yet non-identical, values for the right-handed and left-handed reflective structures indicate the influence of the left- and right-handed structure geometry on the phase synchronization of the electron-field interaction, which may be associated with the specific characteristics of surface wave reflection and propagation.

\onecolumngrid
\section{FIELD DISTRIBUTIONS}
\begin{figure}[!tbh]
  \centering
  \includegraphics[width=\textwidth]{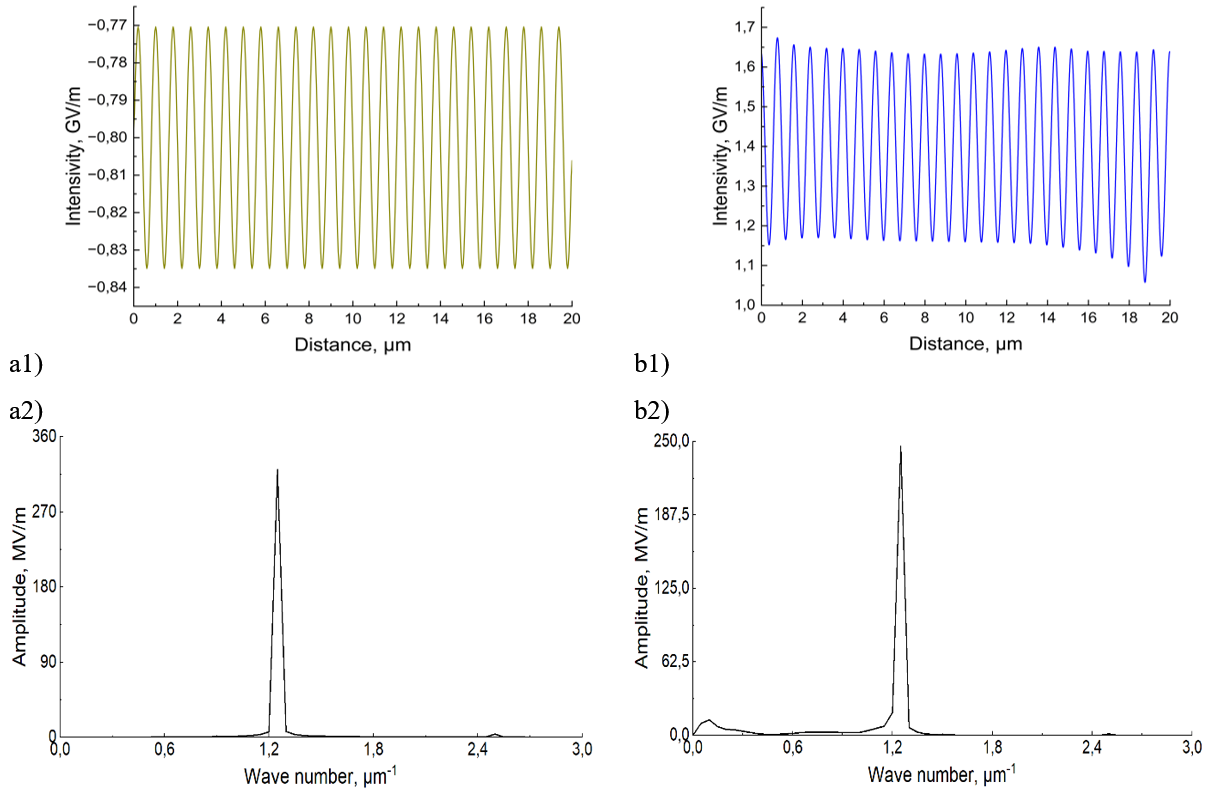}
  \caption{Distribution of the electric field along the surface of a transparent rectangular (a1) and reflective triangular (b1) structure, and the Fourier transform of the respective field data (a2, b2).}
  \label{Fig:09}
\end{figure}

\twocolumngrid

The field distribution (along the electron propagation channel) was analyzed to study the modal composition by applying a Fourier transform to this field. These field distributions were obtained for a transparent rectangular structure and a reflective left-handed structure with a ridge base angle of $\alpha = 20^\circ$ at a specific moment in time. This selection of structures is justified by the fact that this particular reflective structure provides the highest maximum energy gain among all investigated reflective structures.

\Figref{Fig:09} shows the distributions of the longitudinal electric field above the transparent rectangular (a1) and reflective triangular (b1) structures, as well as their spatial spectra (a2 and b2, respectively). The performed Fourier transform of the electric field's spatial distribution revealed a distinct peak for both types of on-chip structures. These peaks correspond to a wavelength close to that of the incident laser radiation. The presence of a sharp, pronounced peak with such a wave number ($k = 1/\lambda \approx 1.2$) indicates that the studied structures with an 800 nm period successfully transform the incident laser radiation into the first spatial harmonic of the excited field. If a relativistic beam ($\beta \approx 1$) is injected into such a field, resonant interaction between this beam and the excited field is ensured. For an electron beam with an initial energy of 10 MeV, as in our simulation, this condition is well satisfied. The high amplitude of the harmonic (relative to the background) indicates efficient field localization along the electron propagation channel and minimal losses to the excitation of parasitic modes.

\section{CONCLUSIONS}
On-chip structures with a rectangular profile and triangular profiles with ridge base angles of $\alpha = 36^\circ$, $30^\circ$, and $20^\circ$ were investigated. The angle selection was based on the availability of ready-made structures from manufacturers. Each triangular structure had both left- and right-handed profile orientations where a left-handed orientation denotes a structure with ridges facing left, and a right-handed orientation denotes a structure with ridges facing right. For all variants, a modified version with a reflective gold coating was additionally considered. The maximum energy gains and accelerating gradients were determined and quantified for all structure classes: transparent and reflective (featuring both rectangular and triangular profiles).

As ridge base angle of transparent structures decreases, the acceleration efficiency is decreased, since the decelerating field propagated from the grooves are increased.

In contrast, for reflective structures, small angles prove to be optimal. In this case, reflection from the ridge facets forms a more concentrated field near the surface, leading to an increase in the accelerating gradient. Furthermore, for reflective structures, the dominant aspect is the interference of reflected waves. Reflective on-chip structures can enhance the acceleration efficiency, particularly for left-handed profiles. Here, the interference between the incident and reflected waves forms a traveling wave that coincides with the direction of the electron beam propagation. For right-handed structures, reflection does not provide as large energy gain, but ensures result stability: the energy gain value depends only weakly on the ridge base angle of the chip structure profile.

The electric field strength of the wave was 1 GV/m. The initial electron energy was $E_{e0}$ = 10 MeV, and the electron beam thickness was 50 nm. The length of all investigated structures was $L$ = 20 $\mu$m. It was demonstrated that left-handed reflective structures provide a maximum energy gain of up to 3 keV and accelerating gradients of up to 150 MeV/m, surpassing the efficiency of their right-handed analogues and nearly all investigated transparent configurations. However, it should be noted that a right-handed transparent on-chip structure with a ridge base angle of $\alpha = 36^\circ$ can provide an energy gain of up to 4.1 keV and an accelerating gradient of up to 205 MeV/m. For a driving wave electric field strength of 1 GV/m, this exceeds previously reported results \cite{Mei2023,Bruckner2024,Yousefi2019}.

By analyzing phase dependencies, optimal injection time windows for a single electron were determined for each structure. It was established that reflective structures, with delay time peaks at 1.4 --1.45 fs, provide a consistently high energy gain within this narrow interval. The reflective rectangular on-chip structure has a broader delay time range (from 0.1 to 0.3 fs), which may allow for the acceleration of a broader beam. It was determined that varying the apex angle and the mirror orientation (left- or right-handed) of the triangular profile can influence the accelerating gradient.

Thus, the results of this work demonstrate that dielectric laser accelerators with a triangular reflective profile, as well as right-handed transparent structures, can provide high accelerating gradients. The obtained results pave the way for the development of compact and efficient next-generation accelerator modules based on such serially produced diffraction gratings.

\begin{acknowledgments}
The study is supported by the National Research Foundation of Ukraine under the program "Excellent Science in Ukraine" (project \# 2023.03/0182).
\end{acknowledgments}

\end{document}